**Logarithmic Tracing and Equilibrium Selection in Asymmetric Coordination Games**

Yali Dong[a], Boyu Zhang[b,*]

[a] School of System Science, Beijing Normal University, 100875, China;

[b] Laboratory of Mathematics and Complex Systems, Ministry of Education, School of Mathematical Sciences, Beijing Normal University, Beijing 100875, China,

[*] Corresponding author: zhangby@bnu.edu.cn

**Abstract:** We study equilibrium selection in asymmetric pure coordination games with finitely many players and actions. A logarithmic regularization defines a tracing-type path of interior equilibria from the uniform strategy profile. We reduce the equilibrium conditions to single-peaked scalar functions and show that, when one associated payoff index is uniquely minimal, the connected equilibrium branch from the uniform profile converges to the corresponding coordinated outcome. With two players, this condition is equivalent to maximizing the sum of square-root payoffs, extending the known two-action result to finitely many actions. A three-player example shows that the square-root rule does not generally extend to more players.

## 1. Introduction

Coordination games often have multiple Nash equilibria. Although each equilibrium is self-enforcing, equilibrium conditions alone do not determine which outcome players will coordinate on. Equilibrium selection procedures address this indeterminacy by specifying how a particular equilibrium is selected. Harsanyi's tracing procedure approaches this problem by following equilibria of auxiliary games from an initial prior toward an equilibrium of the original game (Harsanyi, 1975; Harsanyi and Selten, 1988). Related homotopy methods study equilibrium paths generated by varying a parameter, including paths based on quantal response equilibria (McKelvey and Palfrey, 1995; Turocy, 2005; Zhang, 2016; Zhang, Hofbauer, 2016).

This paper studies a tracing-type selection method for asymmetric pure coordination games with finitely many players and actions. Players receive positive, potentially different payoffs when all choose the same action, and zero payoffs otherwise. We consider a family of games in which each player's material payoff is combined with a logarithmic regularization term. At the initial parameter value, the unique equilibrium is the uniform strategy profile. As the weight on material payoffs increases, we examine the connected component of the equilibrium graph containing that initial equilibrium. This construction is related to logarithmic tracing (Harsanyi, 1973; Zhang, Hofbauer, 2016), but is distinct from Harsanyi's original tracing procedure (Harsanyi, 1975).

Our analysis reduces the equilibrium conditions to a family of single-peaked scalar functions, one for each coordinated outcome. Each function has a maximum that defines an associated payoff index. We show that if one outcome uniquely minimizes this index, the equilibrium component originating at the uniform profile converges to that outcome. For two players, minimizing the index is equivalent to maximizing the sum of square-root payoffs. This extends the two-action result of Zhang and Hofbauer (2016) to any finite number of actions. For three or more players, the square-root rule generally fails, as illustrated by a three-player example.

The paper proceeds as follows. Section 2 introduces the logarithmically regularized game and defines the equilibrium component used for selection. Section 3 establishes the selection result, derives the two-player rule, and gives the three-player example. Section 4 discusses the scope of the criterion and possible extensions.

## 2. Model

Consider an asymmetric pure coordination game with $n \geq 2$ players and $m \geq 2$ actions. Every player has the action set $\{1, \dots, m\}$. If all players choose action $j$, player $i$receives payoff $a_{ij} > 0$, and every other action profile gives every player zero.

Let $x_{ij}$ be the probability that player $i$ chooses action $j$, where $\sum_{j=1}^{m} x_{ij} = 1$. Players randomize independently, so the probability that all players choose action $j$ is $P_j(x) = \prod_{i=1}^{n} x_{ij}$. For $\lambda \geq 0$, player $i$ chooses a mixed strategy to maximize

$$\lambda \sum_{j=1}^{m} a_{ij} P_j(x) + \sum_{j=1}^{m} \log x_{ij} \tag{1}$$

subject to $\sum_{j=1}^{m} x_{ij} = 1$ and $x_{ij} \geq 0$. The logarithmic term is assigned value $-\infty$ at the boundary. The parameter $\lambda$ measures the weight placed on material payoffs relative to logarithmic regularization.[1]

[1] Multiplying (1) by the positive constant $\frac{1}{\lambda+1}$ and writing $\theta = \frac{\lambda}{\lambda+1}$ yields $\theta \sum_{j=1}^{m} a_{ij} P_j(x) + (1-\theta) \sum_{j=1}^{m} \log x_{ij}$, which is the logarithmic game of Harsanyi (1973) in the form used by Zhang and Hofbauer

For each $\lambda$, the objective is strictly concave in player $i$'s own strategy: the material-payoff term is linear in $x_i$, while the logarithmic term is strictly concave. The objective tends to $-\infty$ at the boundary, so each player has a unique interior best response. The best-response map is continuous, theredore Brouwer's fixed-point theorem guarantees an equilibrium. At $\lambda = 0$, each player uniquely maximizes the sum of logarithms by choosing uniformly:

$$x_{ij} = \frac{1}{m} \text{ for all } i, j. \tag{2}$$

This parameterized family defines a tracing-type homotopy: it traces the equilibrium component containing the uniform profile at $\lambda = 0$, and follows equilibria as $\lambda$ increases. We use "racing-type" to describe this construction.

Let $E$ denote the equilibrium graph,

$$E = \{(\lambda, x) \in [0, \infty) \times \Delta_m^n \colon x \text{ is an equilibrium at } \lambda\},$$

and let $C_0$ be the connected component of $E$ containing the uniform equilibrium $(0, \bar{x})$, where $\bar{x}_{ij} = \frac{1}{m}$. Define the limit set of this component by

$$L(C_0) = \{x^* \colon \exists\ (\lambda_r, x^r) \in C_0 \text{ with } \lambda_r \to \infty \text{ and } x^r \to x^*\}.$$

If $L(C_0)$ is a singleton, its element is the outcome selected by the tracing-type procedure. This definition concerns a connected component of the equilibrium graph, and it does not require the equilibria on that component to form a single-valued function of $\lambda$ (Zhang, 2016).

To see that limits along this component are Nash equilibria of the underlying game, let $v_{ij}(x_{-i})$ denote player $i$'s expected material payoff from pure action $j$, and let $\bar{v}_i(x) = \sum_{j=1}^{m} x_{ij} v_{ij}(x_{-i})$ be player $i$'s expected material payoff. The first-order conditions for (1) are

$$\lambda v_{ij}(x_{-i}) + \frac{1}{x_{ij}} = \mu_i,\ j = 1, \dots, m \tag{3}$$

where $\mu_i$ is the multiplier on the probability constraint. Multiplying (3) by $x_{ij}$ and summing over $j$ gives $\mu_i = \lambda \bar{v}_i(x) + m$. Therefore,

$$\lambda \left( v_{ij}(x_{-i}) - \bar{v}_i(x) \right) = m - \frac{1}{x_{ij}}.$$

Since $x_{ij} \leq 1$, the regret from a unilateral deviation satisfies

---

(2016). Since the multiplier is positive, the two specifications have the same equilibria at every $\lambda$.

$$0 \le \max_j v_{ij}(x_{-i}) - \bar{v}_i(x) \le \frac{m-1}{\lambda}. \tag{4}$$

Thus, every element of $L(C_0)$ is a Nash equilibrium of the underlying game. The question is which equilibrium is reached by $C_0$.

### 3. Results

For each action $j$, define $f_j(t) = \frac{t}{\prod_{i=1}^{n}(1+a_{ij}t)}$, $t > 0$, and let $M_j = \max_{t>0} f_j(t)$. The maximum is attained at a unique point, as shown below. Our main result states that if one action uniquely minimizes $M_j$, then the equilibrium component originating at the uniform profile converges to the corresponding pure coordination equilibrium.

**Theorem 1.** Suppose there is a unique action $k$ such that $M_k < M_j$ for every $j \neq k$. Then $\mathcal{L}(\mathcal{C}_0) = \{(e_k, \dots, e_k)\}$, where at this equilibrium every player chooses action $k$.

**Proof outline:** We first express the equilibrium conditions as one scalar equation for each action, with a common level $h$. Each scalar function is single-peaked. Starting from the uniform equilibrium, all actions lie on the increasing branches of their respective functions. Since $M_k$ is uniquely smallest, action $k$ is the first to reach its peak. The central component then continues along the decreasing branch for $k$, while all other actions remain on their increasing branches. As the auxiliary variable $U_k$ tends to infinity, the strategy profile converges to coordination on $k$.

**Proof:** We first characterize the equilibria for $\lambda > 0$. Let $U_j = \lambda P_j(x) = \lambda \prod_{i=1}^{n} x_{ij}$. Multiplying the first-order condition (3) by $x_{ij}$ gives $\mu_i x_{ij} = 1 + a_{ij} U_j$. Summing over $j$ yields

$$\mu_i = m + \sum_{l=1}^{m} a_{il}\, U_l. \tag{5}$$

Using $U_j = \lambda \prod_{i=1}^{n} x_{ij}$ and $x_{ij} = \frac{1+a_{ij}U_j}{\mu_i}$, we obtain

$$f_j(U_j) = h \text{ for every } j,\ h = \frac{\lambda}{\prod_{i=1}^{n} \mu_i} \tag{6}$$

Conversely, suppose positive $U_1, \dots, U_m$ and $h$ satisfy (6). Define

$$\mu_i(U) = m + \sum_{l=1}^{m} a_{il}\, U_l,\ x_{ij}(U) = \frac{1+a_{ij}U_j}{\mu_i(U)}, \text{and } \lambda(U) = h \prod_{i=1}^{n} \mu_i\,(U). \tag{7}$$

These definitions imply $\sum_{j=1}^{m} x_{ij}\,(U) = 1$. Equation (6) then gives $\lambda(U) \prod_{i=1}^{n} x_{ij}(U) = U_j$, so the first-order conditions hold. Strict concavity makes them sufficient for equilibrium. Thus, (6) together with the reconstruction formulas (7), characterizes the equilibrium graph for $\lambda > 0$.

We next show that every $f_j$ is strictly single-peaked. For each $j$,

$$\frac{f_j'(t)}{f_j(t)} = \frac{1}{t}\left(1 - \sum_{i=1}^{n} \frac{a_{ij}t}{1+a_{ij}t}\right). \tag{8}$$

The sum in (8) is strictly increasing from $0$ to $n$. Hence there is a unique $\tau_j > 0$ satisfying

$$\sum_{i=1}^{n} \frac{a_{ij}\tau_j}{1+a_{ij}\tau_j} = 1. \tag{9}$$

It follows that $f_j$ is strictly increasing on $(0, \tau_j)$, strictly decreasing on $(\tau_j, \infty)$, and tends to zero at both endpoints. In particular, $M_j = f_j(\tau_j)$. For each $h \in (0, M_j)$, the equation $f_j(t) = h$ therefore has two roots. Denote the root on the increasing branch by $t_j^-(h)$ and the root on the decreasing branch by $t_j^+(h)$.

We then identify the central component. Since $0 < U_j \leq \lambda$, every equilibrium with sufficiently small positive $\lambda$ lies on the increasing branch of each $f_j$. The assumption $M_k < M_j$ for every $j \neq k$ implies action $k$ is the first to reach its peak as $h$ increases, at $h = M_k$. For $j \neq k$, $h \leq M_k < M_j$, so those coordinates cannot reach their peaks. The component through the uniform equilibrium is therefore described by

$$U_j = t_j^-\left(f_k(U_k)\right),\ j \neq k,\ U_k > 0. \tag{10}$$

For $U_k < \tau_k$, all coordinates lie on their increasing branches; at $U_k = \tau_k$, the two roots for $k$ meet; and for $U_k > \tau_k$, $U_k$ lies on its decreasing branch while the other coordinates remain on their increasing branches. Since $f_k(U_k) \leq M_k < M_j$ for every $j \neq k$, all the roots in (10) are well defined. They vary continuously with $U_k$, and no other coordinate can switch branches. The scalar representation and its inverse $U_j = \lambda P_j(x)$ are continuous on the positive $\lambda$ equilibrium graph. Hence (10), together with its limit at $\lambda = 0$, describes the central component.

Finally, we study the limit of the central component. As $U_k \to \infty$, $f_k(U_k) \to 0$, and (10) implies $U_j \to 0$ for every $j \neq k$. Equation (5) therefore gives $x_{ik} \to 1$ and $x_{ij} \to 0$ for $j \neq k$. Consequently, $P_k(x) \to 1$ and $\lambda = \frac{U_k}{P_k(x)} \to \infty$. Conversely, the reconstruction formulas (7) extend continuously to $U_k = 0$, with $\lambda(0) = 0$, and are bounded on every bounded interval of $U_k$. Hence any sequence on $C_0$ with $\lambda \to \infty$ must satisfy $U_k \to \infty$. It follows that $\mathcal{L}(C_0) = \{(e_k, \ldots, e_k)\}$. This proves the theorem. □

The theorem gives a selection criterion for the tracing-type component when the minimum of the payoff indices is unique. If several actions share the minimum, the argument does not determine whether the component has a unique limit.

For two players, the index has a simpler form. For each action $j$,

$$\frac{1}{M_j} = \min_{t>0} \frac{(1+a_{1j}t)(1+a_{2j}t)}{t} = (\sqrt{a_{1j}} + \sqrt{a_{2j}})^2. \tag{11}$$

The last equality follows by minimizing $\frac{1}{t} + a_{1j}a_{2j}t$. Thus, minimizing $M_j$ is equivalent to maximizing the sum of square-root payoffs.

**Corollary 1.** For two players and any finite number of actions, if one action uniquely maximizes $\sqrt{a_{1j}} + \sqrt{a_{2j}}$, then the component originating at the uniform profile converges to the pure coordination equilibrium on that action.

This extends the two-action square-root selection rule of Zhang and Hofbauer (2016). It is worth noting that the rule does not generally extend to three or more players. For example, let the payoff vectors at actions $A$ and $B$ be $a = (16, 1, 1)$, $b = \left(\frac{441}{100}, \frac{441}{100}, \frac{441}{100}\right)$. The square-root sums favor B since $\sum_{i=1}^{3} \sqrt{b_i} = 6.3 > 6 = \sum_{i=1}^{3} \sqrt{a_i}$. However, the indices $M_A$ and $M_B$ favor $A$. For $A$, $\frac{1}{M_A} = \min_{t>0} \left(\frac{1}{t} + 18 + 33t + 16t^2\right)$, where the unique minimizer is $\tau_A = \frac{\sqrt{129}-1}{64}$. Substitution yields $M_A \approx 0.03340$. For $B$, the three payoffs are equal, and direct maximization gives $M_B \approx 0.03359$. Since $M_A < M_B$, Theorem 1 selects $A$. This example shows that the two-player square-root rule does not generally characterize selection with three players.

## 4. Discussion

This paper gives a selection criterion for asymmetric pure coordination games under logarithmic regularization. The equilibrium graph is traced from the uniform profile, where material payoffs have no weight. When one action uniquely minimizes the index $M_j$, the connected component $C_0$ converges to coordination on that action. For two players, minimizing $M_j$ is equivalent to maximizing the sum of square-root payoffs. With three or more players this simplification generally fails, and $M_j$ rather than the square-root sum determines selection under the unique-minimizer condition.

The index $M_j$ is related to, but distinct from, several established equilibrium selection criteria. In asymmetric pure coordination games, an outcome that Pareto dominates every other coordinated outcome necessarily has a strictly smaller index $M_j$. The same implication holds for a majority dominant outcome in Zhang (2016), which is an equilibrium selection criterion for a broad class of quantal response functions. In asymmetric pure coordination games, majority dominance is equivalent to strict Pareto dominance, which in turn implies that $M_j$ is uniquely minimized at $j$. By contrast, the criterion is nevertheless distinct from aggregate-payoff maximization and risk dominance. In the two-player, two-action case, the square-root ranking can differ from both the ranking by total payoffs and the ranking by the product of deviation losses, the latter characterizing risk dominance in this setting (Zhang and Hofbauer 2016).

Our results rely on the structure of the game and the logarithmic regularization. In particular, the scalar representation used in the proof exploits the fact that only coordinated action profiles yield positive payoffs. The analysis therefore does not yield a comparable rule for general coordination games or for other quantal response functions such as logit response (Zhang and Hofbauer 2016). The main theorem also requires a unique minimizer of $M_j$. When several actions attain the minimum, the argument does not show whether $C_0$ has a unique limit.

Several extensions merit further study. One is to examine whether the square-root ranking provides a useful first-order approximation for two-player games with more than two actions under more general quantal response functions, and to identify conditions under which a local approximation determines the global equilibrium branch (Zhang and Hofbauer, 2016). Another is to characterize the limiting equilibria of logit and other quantal response correspondences in multi-action coordination games (McKelvey and Palfrey, 1995; Turocy, 2005; Zhang, 2016). Finally, it would be useful to analyze how equilibrium selection changes under payoff transformations or the addition of dominated strategies (Zhang, 2016), and to compare logarithmic tracing with alternative selection methods, including linear tracing procedures (Harsanyi, 1975; Cao and Dang, 2022), homotopy methods (Turocy, 2005), and evolutionary dynamics (Zhang and Hofbauer, 2015).

**Acknowledgements**: Yali Dong acknowledges support from the National Natural Science Foundation of China (Grant No 72573024). Boyu Zhang acknowledges support from the National Natural Science Foundation of China (Grant Nos. 72131003 and 72671038) and the Beijing Natural Science Foundation (Grant No. Z220001). The authors acknowledge the use of ChatGPT to assist in polishing the language. All content has been reviewed and edited under the authors' oversight and control. No artificial intelligence tools were used for idea generation or development of research findings.

**Conflict of interest:** The author declares no competing interests.

**Author Contributions:** Conceptualization, B.Z.; methodology, B.Z.; formal analysis, Y.D., B.Z.; writing - original draft, Y.D., B.Z.; writing - review & editing, Y.D., B.Z.; funding acquisition Y.D., B.Z.